\documentclass[10pt,twocolumn,graphicx,psfig,amssymb,amsthm,amsmath,conference]{IEEEtran}
\IEEEoverridecommandlockouts
\usepackage{amsmath,amssymb,amsfonts}
\usepackage{mathtools, cuted}
\usepackage{algorithmic}
\usepackage{graphicx}
\usepackage{steinmetz}
\usepackage{textcomp}
\usepackage[utf8]{inputenc}
\usepackage{cite}
\usepackage{lettrine}
\usepackage{mathtools}
\usepackage[most]{tcolorbox}
\usepackage{dblfloatfix} 
\usepackage{booktabs,xcolor,siunitx}
\definecolor{lightgray}{gray}{0.9}
\usepackage{multicol, blindtext}
\usepackage{multirow}
\usepackage{float}
\usepackage{epstopdf}
\usepackage{siunitx}
\usepackage{csquotes}
\usepackage{subfigure}
\usepackage{romannum}
\usepackage{etoolbox}
\newtheorem{remark}{Remark}
\begin{document} 
\title{\LARGE Mixed Dual-Hop IRS-Assisted FSO-RF Communication System with H-ARQ Protocols\vspace{-0.8em}}
%\author {Author 1, Author 2, Author 3, and\hspace*{-0.09em} Author 4 %\vspace*{-3em}
\author {Gyan Deep Verma, Aashish Mathur, Yun Ai, and\hspace*{-0.09em} Michael Cheffena \vspace*{-1em}
%\author {Gyan Deep Verma, Aashish Mathur, $Member,\hspace*{-0.09em} IEEE$, and\hspace*{-0.09em} Yun Ai, $Member,\hspace*{-0.09em}IEEE$  \vspace*{-1em}
		%\thanks{Gyan Deep Verma and Aashish Mathur are with the Department of Electrical
		%	Engineering, Indian Institute of Technology Jodhpur, Jodhpur,
		%	342037, India (e-mails: \{verma.11, aashishmathur\}@iitj.ac.in). 
			
		%	Yun Ai is with the Norwegian University of
			%Science and Technology (e-mails: yun.ai@ntnu.no).
			
		%}
	}
\maketitle 

\begin{abstract}
Intelligent reflecting surface (IRS) is an emerging key technology for the fifth-generation (5G) and beyond wireless communication systems to provide more robust and reliable communication links. In this paper, we propose a mixed dual-hop free-space optical (FSO)-radio frequency (RF) communication system that serves the end user via a decode-and-forward (DF) relay employing hybrid automatic repeat request (H-ARQ) protocols on both hops. Novel closed-form expressions of the probability density function (PDF) and cumulative density function (CDF) of the equivalent end-to-end signal-to-noise ratio (SNR) are computed for the considered system. Utilizing the obtained statistics functions, we derive the outage probability (OP) and packet error rate (PER) of the proposed system by considering generalized detection techniques on the source-to-relay (S-R) link with H-ARQ protocol and IRS having phase error. We obtain useful insights into the system performance through the asymptotic analysis which aids to compute the diversity gain. The derived analytical results are validated using Monte Carlo simulation. 
\end{abstract}

\vspace{-0.9em}
\section{Introduction}
\vspace{-0.3em}
\lettrine[findent=2pt]{\textbf{I}}{N} recent decade, free-space optical (FSO) communication systems have received significant
attention in academic and industrial research. FSO systems feature a license-free band, higher bandwidth, no electromagnetic interference (EMI), and a more secure communication link compared to traditional radio frequency (RF) communication systems due to the inherent properties of optical laser beams \cite{khaligi:01}.  However, the atmospheric turbulence (AT) and pointing errors (PEs) may lead to severe degradation in intensity, phase, and polarization of the received signal, especially for the long distance communications \cite{khaligi:01, yang:02,verma:01}.  
	
Various methods have been developed by researchers to overcome the effects of AT and PEs effectively, one of which is to use co-operative communication \cite{yang:01}. A mixed dual-hop FSO-RF wireless communication system has been a promising approach for leveraging both wireless techniques to increase capacity, reduce power consumption, and extend wireless coverage by compensating the deteriorating effects of AT and PEs. The decode-and-forward (DF) relay is an intermediate relay node that is used to connect both hops. The DF relay node's objective is to decode the incoming signal and re-transmit it to the next hop.
To further improve the performance of the communication systems, hybrid automatic repeat request (H-ARQ) techniques can be utilized, which combines forward error correction (FEC) with re-transmission request protocols \cite{3gpp:01}, \cite{mathur:01}. The approximate packet error rate (PER) of an RF communication system using H-ARQ with chase combining (CC) protocol was obtained in \cite{ge:01}, \cite{xi:01}.  The authors in \cite{verma:02}, computed the outage probability (OP) and throughput performance of FSO communication systems employing various H-ARQ protocols under the combined influence of Gamma-Gamma (GG) distributed AT and PEs.   

The use of intelligent reflecting surface (IRS) has been proposed very recently in literature to improve the RF system performance. The IRS is made up of a large number of reflecting active antenna plates with a micro-controller, which enables to control amplitude, frequency, phase, and polarization of the incident electromagnetic signals. Moreover, it does not require much signal processing, such as encoding, decoding, and amplification \cite{bjornson:01}. The IRS or meta-surfaces have been extensively explored in the literature, for their merits such as, capability of the greater coverage, improved spectrum and energy efficiency \cite{yang:01}, \cite{sikri:01, ai:01, jamali:01}. In \cite{yang:01}, the authors reported OP and BER performance for an IRS-aided dual-hop FSO-RF communication system. An IRS-assisted dual-hop mixed FSO-RF communication system was recently proposed in \cite{sikri:01} where the OP and BER performance under the influence of co-channel interference (CCI) were evaluated. The authors in \cite{ai:01} performed OP analysis for an RF communication system using the H-ARQ with CC protocol over an IRS assisted with perfect phases estimation. However, the effect of phase errors in IRS was not considered in the aforementioned works.  

Motivated by the recent studies on mixed FSO-RF communication systems using IRS and the benefits of H-ARQ protocols to improve the reliability of communication systems, we study the performance of an IRS-assisted mixed dual-hop FSO-RF communication system with H-ARQ protocols in this work. To the best of the authors' knowledge, such a cascaded system employing H-ARQ protocols over both links has not been investigated in literature. The following are the key contributions of our research work:
\begin{enumerate}
		\item We derive closed-form expression of the probability density function (PDF) and cumulative distribution function (CDF) of the accumulated signal-to-noise ratio (SNR) with imperfect IRS phase and H-ARQ protocol. Furthermore, using these expressions, we  calculate the theoretic information OP and PER for the considered system.
\item We analyze the impact of various system parameters such as the number of IRS reflecting plates, number of transmission rounds, von Mises distribution concentration parameter (phase
estimation accuracy of IRS), and packet length on the OP and PER performance which provide significant design insights into the considered system.
\item Useful insights into the system performance are obtained through the asymptotic expression of the OP.
\end{enumerate}
\vspace{-0.9em}
 \section{System Description}
We consider a cascaded dual-hop FSO-RF communication system where H-ARQ technique is employed on both the links. We assume that a source node ($S$) communicates with a destination node ($D$) via intermediary DF based relay ($R$). Due to obstructions caused by high-rise buildings and other impediments, it is justified that there is no direct path between source-to-destination ($S$-$D$) and relay-to-destination ($R$-$D$) \cite{yang:01}. Hence, IRS is utilized on the RF link ($R$-$D$).
\begin{figure}[t]
	\vspace{-1.0em}
\centering
\includegraphics[width=0.98\linewidth]{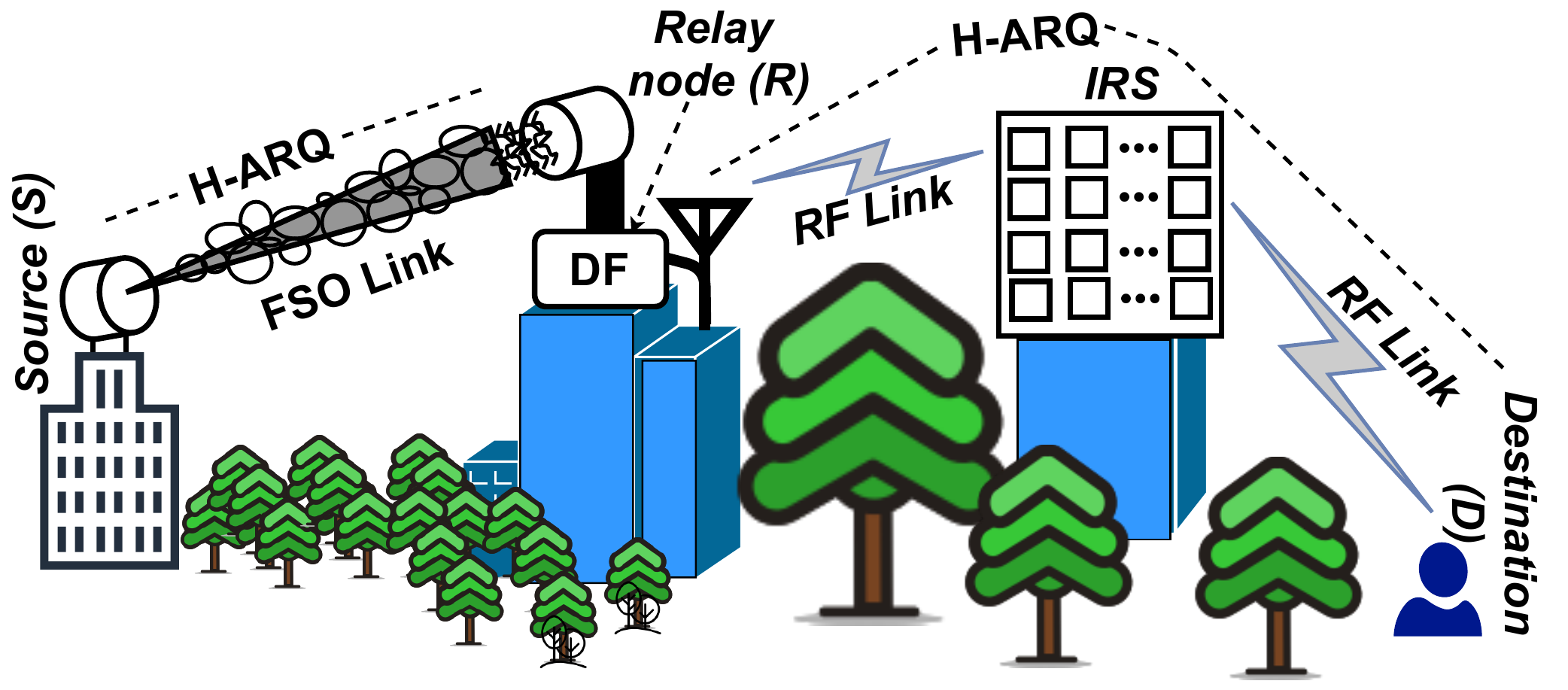}
	\vspace{-0.6em}
										\caption{Mixed dual-hop FSO-RF communication system model with IRS and H-ARQ protocol.}
											\vspace{-1.8em}
			\end{figure}
The $S$-$R$ link is a FSO link which is subjected to GG AT and PEs. For the FSO signal reception, both the heterodyne detection (HD) and intensity modulation and direct detection (IM/DD) techniques can be used. Rician and Rayleigh distributions are used on the RF hop between $R$-IRS and IRS-$D$, respectively, while von Mises distribution describes the phase shift errors of the IRS \cite{badiu:01}.         
\subsection{FSO Transmission Link}	
On the FSO communication link, we assume a single transmitter (Tx) and receiver (Rx) and use H-ARQ with CC protocol to enhance the system performance. The baseband electrical signal is represented after optical-to-electrical (O-E) conversion as follows:
\vspace{-0.9em}
\begin{equation}
y_{SR}= \eta_{e}s_1 h+w_{R},
\label{system_model}
\vspace{-0.5em}
\end{equation}
where $\eta_{e}$ is the effective O-E conversion coefficient and the transmitted optical signal is $s_1$. The FSO channel gain is represented as $h$=$h_{l}h_{a}h_{p}$, where $h_{l}$ denotes path loss which arises due to dust, rain, fog, snow, and other atmospheric impairments \cite{ghassemlooy:01}, $h_a$ signifies AT, which is caused by inhomogeneities present along the optical beam path, and $h_p$ indicates PE, which is caused by thermal expansion, earthquakes, and other factors \cite{yang:02}. It is assumed that the transmitted power is $\mathbb{E}\left [s_1\right ]$=$P_1$, where $\mathbb{E}\left [.\right]$ is the expectation operator, $w_R \sim C\mathcal{N}(0,\sigma^{2}_{R})$ denotes the complex additive white
Gaussian noise (AWGN) with zero mean and variance $\sigma^{2}_{R}$ at the relay node \cite{verma:01}. The composite PDF of the received irradiance at the relay in the $q^{th}$ round is expressed as \cite[Eq.~(8)]{bhatnagar:01},
\vspace{-0.7em}
\begin{align}
     f_{h_{SR_q}}(h_{q})=\frac{a b \xi^2}{{A}_{0}h_{q}\Gamma (a)\Gamma (b)}G_{1,3}^{3,0}\Big (\frac{a b h_{q}}{A_{0}}\Big | {{\xi ^2+1} \atop {\xi^2,a,b}}\Big). \label{composite_PDF}
\end{align}
In (\ref{composite_PDF}),  $G^{s,t}_{u,v}\left(z\left|{\begin{array}{c}c_{1},\dots, c_{u}\\d_{1},\dots,d_{v}\end{array}}\right.\right)$ represents the Meijer G-function \cite[Eq. (9.301)]{gradshteyn:01} and  $\Gamma(.)$ denotes the Gamma function \cite[Eq. (8.31)]{gradshteyn:01}. The system parameters $a$, $b$, and $\xi^{2}$ are discussed in depth in \cite{bhatnagar:01}. After $N_1$ independent and identical H-ARQ transmission rounds over the $S$-$R$ link, the random variable (RV) for the accumulated SNR is $Z_{SR}$=$\sum_{q=1}^{N_1}\gamma_{SR_q}$, where $\gamma_{SR_q}$=$(\eta_{e}h_q)^r/\sigma^{2}_{R}$ represents the instantaneous SNR on the $S$-$R$ link in the $q^{th}$ round and $r$=$\left \{1,2  \right\}$ denotes HD and IM/DD, respectively. Further, using \cite[Eq. (26)]{bhatnagar:01}, the PDF of the RV $Z_{SR}$ is written as 
\vspace{-1.5mm}
\begin{equation}f_{Z_{SR}}(z)=\hspace{-2em}\sum _{{l_1+l_2+l_3}=N_1\atop {0\le l_1,l_2,l_3\le N_1}}\hspace{-0.7em}\left({N_1}\atop {l_1,l_2,l_3}\right)\sum ^\infty _{n=0}\frac{D_{n}(l_1,l_2,l_3)z^{\frac{E}{r}-1}}{\Gamma\left ( \frac{E}{r} \right )\bar{\gamma}_{SR_r}^{\frac{E}{r}}}.\vspace{-0.7em}\label{trans pdf}\end{equation} 
 In (\ref{trans pdf}), the parameters $D_{n}(l_1,l_2,l_3)$=$\tilde{X_{0}}^{l_1}(\tilde{Y_{n}}^{(l_2)}*\tilde{Z_{n}}^{(l_3)})$ are defined in  \cite[Eq.~(13)]{bhatnagar:01}, $\bar{\gamma}_{SR_r}$=$(\eta_{e}E[h_q])^r/\sigma^{2}_{R}$, $E\left[h_q\right]$=$(A_0\xi^2)/(\xi^2+1)$, $
E$=$n+l_1\xi^{2}+l_{2}a +l_{3}b$, $*$ denotes convolution, and $\tilde{Y_{n}}^{(l_2)}$ shows that $\tilde{Y_{n}}$ is convoluted $(l_2-1)$ times with itself. Moreover, by integrating (\ref{trans pdf}) with respect to $z$, the CDF of the RV $Z_{SR}$ is derived as
\vspace{-1.5mm}
\begin{equation} F_{Z_{SR}}(z)=\hspace{-2em}\sum _{{l_1+l_2+l_3}=N_1\atop {0\le l_1,l_2,l_3\le N_1}}\hspace{-0.9em}\left({N_1}\atop {l_1,l_2,l_3}\right)\sum ^\infty _{n=0}\frac{D_{n}(l_1,l_2,l_3)z^{\frac{E}{r}}}{\Gamma\left ( \frac{E}{r} \right )\left ( \frac{E}{r} \right )\bar{\gamma}_{SR_r}^{\frac{E}{r}}}.\vspace{-0.7em}\label{CDF_SNR}\end{equation}
\subsection{RF Transmission Link}
We assume that the $R$-$D$ link is an RF link that utilizes IRS with $M$ identical antenna reflector plates that serve as an intermediary node between R and D (spaces between two plates are maintained at least half wavelength). The RF signal after reflection from the IRS and received at $D$ is written as follows \cite{badiu:01}:
\vspace{-1.7mm}
\begin{equation} y_{RD} = M\sqrt {\gamma_{0}}\left[{ \sum _{i=1}^{M} I_{1_i} I_{2_i} I_{3_i} e^{j\Theta_i}}\right] s_{2}+ w_{D}, \vspace{-0.6em}\end{equation}
 where $s_2$ is the transmitted symbol from $R$ to $D$ and $\mathbb{E}\left [s_2\right ]$=$1$, $\gamma_{0}$ represents the average SNR when single reflecting plate is present, $I_{1_i}$=$(d_{1})^{-\nu /2}\alpha_{i}\exp(-j\theta_{i})$ and $I_{2_i}$=$(d_{2})^{-\nu /2}\beta_{i}\exp(-j\psi_{i})$ are the channel gains of the $R$-RIS and RIS-$D$ links, and $\nu$ is the path loss exponent. Further, $d_1$, $\theta_{i}$, and $\alpha_{i}$ are the distance, phase shift, and the Rician fading channel on the $R$-IRS link, respectively \cite{badiu:01}. Likewise, $d_2$, $\psi_{i}$, and $\beta_{i}$ are the distance, phase shift, and the Rayleigh fading channel on the IRS-$R$ link, respectively \cite{badiu:01}. $I_{3_i}$=$\delta_{i}\exp(j\phi_{i})$ is the reflection coefficient induced by the $i^{th}$ IRS reflector, $\delta_{i}$=$1$, $\forall_i$ and $w_D \sim C\mathcal{N}(0,\sigma^{2}_{D})$ denotes the complex AWGN with zero mean and $\sigma^{2}_{D}$ variance at node $D$.
The authors of \cite{sikri:01}, \cite{ai:01}, \cite{kong:01} assumed that the IRS estimates perfect phase shift  $\phi_{i}$=$\theta_{i}$+$\psi_{i}$ which nullifies the total phase shift that results in maximizing the SNR at $D$. However, in practical scenarios the phase shift induced by the channel is difficult to perfectly estimate \cite{badiu:01}. Therefore, in this work, we incorporate the phase deviation defined as  $\theta_{i}$+$\psi_{i}$-$\phi_{i}$=$\Theta_{i}$ in our analysis, where $\Theta_{i}$ is randomly distributed
over $[-\pi, \pi)$ and is modeled by the circular  distribution.
The combined PDF of the $R$-IRS and IRS-$D$ channels for the $j^{th}$ transmission round after incorporating phase shift errors is comparable to a direct channel with Nakagami distribution and is given by \cite[Eq. (11-12)]{badiu:01}
\vspace{-1.7mm}
\begin{equation} 
f_{|I_{RD_j}|}(x) = \frac {2m^{m}}{\Gamma (m)\mu ^{2m}} x^{2m-1}\exp \left ({-\frac {m}{\mu ^{2}}x^{2}}\right),\vspace{-0.55em}\label{RF_fading}
\end{equation}
where $|I_{RD_j}|$=$\frac{1}{M}\sum _{i=1}^{M} |I_{1_i}| |I_{2_i}| |I_{3_i}|  \exp({j\Theta_i})$ , $\mu$ is the common mean of $M$ complex variables, and $m$ is the shaping parameter. The parameter $m$ depends on $M$, $I_\vartheta(k)$, $E[I_{1_i}]$, and $E[I_{2_i}]$, where $I_{\vartheta}(k)$ represents modified Bessel function of the first kind of order $\vartheta$, and $k$ is concentration parameter of  von Mises distribution \cite[Eq. (12)]{badiu:01}. The instantaneous SNR of the $j^{th}$ transmission round at $D$ is $\gamma_{RD_j}$=$M\gamma_0|I^{2}_{RD_j}|$. Now, using (\ref{RF_fading}), the PDF of the instantaneous SNR on the $R$-$D$ link on the $j^{th}$ H-ARQ round is computed as follows:
\vspace{-1.8mm}
\begin{equation} f_{\gamma_{RD_j}}(\gamma_{RD_j}) = \frac {m^{m}}{\Gamma (m)\bar{\gamma}_{RD}^{m}} \gamma_{RD_j}^{m-1}\exp \left ({-\frac {m}{\bar {\gamma}_{RD} }\gamma_{RD_j}}\right),\vspace{-0.6em}\label{RF_SNR}\end{equation}
where $\bar{\gamma}_{RD}$=$M^2\gamma_{0}$. Let $Z_{RD}$=$\sum_{j=1}^{N_2}\gamma_{RD_j}$ represent the accumulated instantaneous SNR after $N_{2}$ independent and identical H-ARQ transmission rounds at node $D$. Utilizing (\ref{RF_SNR}), the moment generating function (MGF) of the RV $Z_{RD}$ is calculated as 
\vspace{-2mm}
\begin{equation}
{\cal M}_{Z_{RD}}\left(s \right)=\frac{m^{N_2}\Gamma(m)^{N_{2}}}{{\left (  s+\frac{m}{\bar{\gamma}_{RD}}\right )^{mN_{2}}}}.\vspace{-0.7em}\label{MGF_result_RF}
\end{equation}
Taking the inverse Laplace transform of (\ref{MGF_result_RF}) and after some algebra, the PDF of the RV $Z_{RD}$ is obtained as
\vspace{-2mm}
\begin{align}
f_{Z_{RD}}(t)=\left ( \frac{mt}{\bar{\gamma}_{RD}} \right )^{N_{2}}\frac{e^{\frac{mt}{\bar{\gamma}_{RD}}}}{(N_{2}m-1)!}.\vspace{-0.7em}\label{RF_PDF}
\end{align}
Integrating (\ref{RF_PDF}), the CDF of $Z_{RD}$ is derived as 
\vspace{-1.5mm}
\begin{align}
F_{Z_{RD}}(t)=\frac{\bar{\gamma}_{RD}}{m}\frac{\Upsilon\left (\frac{mt}{\bar{\gamma}_{RD}},N_{2}m-1\right )}{(N_{2}m+1)!},\vspace{-0.7em}\label{RF_CDF}
\end{align}
where $\Upsilon(. , .)$ is the lower incomplete Gamma  function \cite[Eq. (8.350)]{gradshteyn:01}. 
\vspace{-0.50em}
\section{Performance Analysis}
 \subsection{Equivalent end-to-end SNR}
The CDF of the equivalent end-to-end SNR for the aforementioned communication system is given by \cite[Eq.~(7)]{gupta:01}
\vspace{-5mm}
\begin{align}
  F_{\gamma_{eq}}(\gamma)&=\text {Pr}[\min (Z_{SR},Z_{RD})<\gamma]
  \nonumber\\&=\text {Pr}(Z_{SR}<\gamma ) + \text {Pr}(Z_{RD}<\gamma ) 
  \nonumber\\&~~~-\text {Pr}(Z_{SR}<\gamma )\text {Pr}(Z_{RD}<\gamma).\vspace{-0.7em} \label{equivalent_CDF} 
\end{align} 
The PDF of the equivalent SNR is obtained by simply differentiating (\ref{equivalent_CDF}) as follows:
\vspace{-1.5mm}
\begin{align}
f_{\gamma_{eq}}(\gamma)&=f_{Z_{SR}}(\gamma)+f_{Z_{RD}}(\gamma)-f_{Z_{SR}}(\gamma)F_{Z_{RD}}(\gamma)\nonumber\\&~~- F_{Z_{SR}}(\gamma)f_{Z_{RD}}(\gamma).
\label{eq_PDF}\end{align}
The PDF of the equivalent SNR for the considered system is thus obtained by substituting (\ref{trans pdf}), (\ref{CDF_SNR}), (\ref{RF_PDF}), and (\ref{RF_CDF}) into (\ref{eq_PDF}).
\subsection{OP Analysis}
An information-theoretic outage occurs when the mutual information of the $N$ H-ARQ transmission rounds falls below the required threshold information rate ($R$). Hence, the OP of the considered communication system is given by   
\vspace{-1.5mm}
\begin{equation}
P_{o}={\rm Pr}\{I_{1}\leq R, \ldots, I_{N}\leq R\},    
\end{equation}
where $I_{p}$, $1\leq p\leq N$, denotes the  instantaneous mutual information in the $p^{th}$ round.
In CC protocol, the mutual information is calculated using the accumulated SNR at the Rx after $N$ transmission rounds as  $I_{CC}(N)=\log_2\left(1+\sum_{p=1}^{N}\gamma_{p}\right)$. Hence, the OP for H-ARQ with CC protocol after $N$ rounds can be computed as
\vspace{-1.5mm}
 \begin{align}
     P_{o}={\rm Pr}\bigg \{\hspace{-0.2em}\log_2\bigg(\hspace{-0.1em}1+\hspace{-0.3em}\sum_{p=1}^{N}\gamma_{p}\bigg)\leq R \bigg\}\hspace{-0.1em}=\hspace{-0.1em}{\rm Pr}\bigg \{\hspace{-0.3em}\sum_{p=1}^{N}\gamma_{p}\leq (2^{R}-1) \hspace{-0.1em}\bigg\}.\vspace{-0.5em}\label{CC}
 \end{align}
Using (\ref{equivalent_CDF}) and (\ref{CC}), the end-to-end OP for the considered system is obtained as 
\begin{equation}
    P_{o,CC}=P_{o,SR}+P_{o,RD}-P_{o,SR}P_{o,RD},\label{P_o_cc}
\end{equation}
where $P_{o,SR}$ and $P_{o,RD}$ are the OP for H-ARQ with CC protocol on $S$-$R$ and $R$-$D$ link, respectively. Further, on substituting (\ref{CDF_SNR}) and (\ref{RF_CDF}) into (\ref{P_o_cc}), the OP of the considered system is derived as   
\vspace{-1.5mm}
\begin{align} &P_{o,CC}=\hspace{-5mm}\sum _{{l_1+l_2+l_3}=N_1\atop {0\le l_1,l_2,l_3\le N_1}}\hspace{-0.9em}\left({N_1}\atop {l_1,l_2,l_3}\right)\sum ^\infty _{n=0}\frac{D_{n}(l_1,l_2,l_3)(2^R-1)^{\frac{E}{r}}}{\Gamma\left ( \frac{E}{r} \right )\left ( \frac{E}{r} \right )\bar{\gamma}_{SR_r}^{\frac{E}{r}}}\nonumber 
\\&+\frac{\bar{\gamma}_{RD}}{m}\frac{\Upsilon\left (\frac{m(2^R-1)}{\bar{\gamma}_{RD}},N_{2}m+1\right )}{(N_{2}m-1)!}-\frac{\Upsilon\left (\frac{m(2^R-1)}{\bar{\gamma}_{RD}},N_{2}m+1\right )}{(N_{2}m-1)!}\nonumber\\& \times \frac{\bar{\gamma}_{RD}}{m}\hspace{-5mm}\sum _{{l_1+l_2+l_3}=N_1\atop {0\le l_1,l_2,l_3\le N_1}}\hspace{-0.9em}\left({N_1}\atop {l_1,l_2,l_3}\right)\sum ^\infty _{n=0}\frac{D_{n}(l_1,l_2,l_3)(2^R-1)^{\frac{E}{r}}}{\Gamma\left ( \frac{E}{r} \right )\left ( \frac{E}{r} \right )\bar{\gamma}_{SR_r}^{\frac{E}{r}}}.\vspace{-1.8em}
\label{equivalent_outage_final}
\end{align}
\subsection{Average Packet Error Rate Analysis}
In this section, we will derive the PER for the considered dual-hop FSO-RF communication system. The PER of H-ARQ with CC protocol for a long packet over slow block fading channel after the $p^{th}$ transmission round can be written as \cite[Eq. (2)]{ge:01}
\vspace{-2.5mm}
\begin{equation}
\mathrm{PER}\!=\!\int\limits_{0}^{\infty}\!\cdots\!\int\limits_{0}^{\infty}Q_{1}\!\cdots\! Q_{p}f(\gamma_1)\cdots f(\gamma_p)\, d\gamma_1\cdots d\gamma_{p},
\vspace{-0.7em}\end{equation}
where $Q_{i}$, $i\in{(1,2,\cdot\cdot\cdot,p)}$ is the instantaneous PER under AWGN, i.e.,  $Q_{i}$=$g(\gamma)$ shows a steep waterfall characteristic with threshold $T_0$, $g(\gamma/\gamma\leq T_{0})\approx 1$ and $g(\gamma/\gamma>T_{0})\approx0$. Further, $T_{0}$ is computed as $T_{0}$=$\int_{0}^{\infty}g(\gamma
)d\gamma$, hence the approximated PER is given by \cite[Eq. (3)]{ge:01},\cite[Eq. (4-5)]{xi:01}
\vspace{-1.5mm}
 \begin{equation}
  \mathrm{PER}\!\approx\!\int_{0}^{T_0} f_{\gamma_{eq}}(\gamma)d\gamma=J_1+J_2-J_3-J_4
  \label{PER_integ},\end{equation}
where $J_1$=$\int_{0}^{T_0} f_{Z_{SR}}(\gamma)d\gamma,~ J_2$=$\int_{0}^{T_0} f_{Z_{RD}}(\gamma) d\gamma,~ J_3$=$\int_{0}^{T_0} f_{Z_{SR}}(\gamma)\\ \times F_{Z_{RD}}(\gamma) d\gamma$, and $J_4$=$\int_{0}^{T_0} F_{Z_{SR}}(\gamma)f_{Z_{RD}}(\gamma) d\gamma$. $J_1$ and $J_2$ are calculated by using (\ref{CDF_SNR}) and (\ref{RF_CDF}), respectively. Moreover, $J_3$ is derived with help of (\ref{trans pdf}) and (\ref{RF_CDF}) as follows:
\vspace{-1.5mm}
\begin{align}
    J_{3}&= \hspace{-4mm}\sum _{{l_1+l_2+l_3}=N_1\atop {0\le l_1,l_2,l_3\le N_1}}\hspace{-0.9em}\left({N_1}\atop {l_1,l_2,l_3}\right)\sum ^\infty _{n=0}\frac{D_{n}(l_1,l_2,l_3)}{\Gamma\left ( \frac{E}{r} \right )\bar{\gamma}_{SR_r}^{\frac{E}{r}}}\frac{\bar{\gamma}_{RD}}{m} \nonumber\\&\times\int_{0}^{T_0}\gamma^{\frac{E}{r}-1}\frac{\Upsilon\left (\frac{m\gamma}{\bar{\gamma}_{RD}},N_{2}m+1\right )}{(N_{2}m-1)!}d\gamma.\label{I3}
\end{align}
\begin{figure*}[!t]
\vspace{-3em}
\begin{align}
&\mathrm{PER}\approx\hspace{-5mm}\sum _{{l_1+l_2+l_3}=N_1\atop {0\le l_1,l_2,l_3\le N_1}}\hspace{-0.9em}\left({N_1}\atop {l_1,l_2,l_3}\right)\sum ^\infty _{n=0}\frac{D_{n}(l_1,l_2,l_3)(T_{0})^{\frac{E}{r}}}{\Gamma\left ( \frac{E}{r} \right )\left ( \frac{E}{r} \right )\bar{\gamma}_{SR_r}^{\frac{E}{r}}} +\frac{\bar{\gamma}_{RD}}{m}\frac{\Upsilon\left (\frac{mT_0}{\bar{\gamma}_{RD}},N_{2}m-1\right )}{(N_{2}m-1)!} -\hspace{-6mm}\sum _{{l_1+l_2+l_3}=N_1\atop {0\le l_1,l_2,l_3\le N_1}}\hspace{-0.7em}\left({N_1}\atop {l_1,l_2,l_3}\right)\sum ^\infty _{n=0}\frac{D_{n}(l_1,l_2,l_3)}{\Gamma\left ( \frac{E}{r} \right )\bar{\gamma}_{SR_r}^{\frac{E}{r}}}  \left ( \frac{m}{\bar{\gamma}_{RD}}\right )^{mN_2}\nonumber\\& \times B(1,u)~ _2F_2\left(mN_2+1,u;mN_2+2,u+1;\frac{T_0m}{\bar{\gamma}_{RD}}\right)- \hspace{-5mm}\sum _{{l_1+l_2+l_3}=N_1\atop {0\le l_1,l_2,l_3\le N_1}}\hspace{-0.9em}\left({N_1}\atop {l_1,l_2,l_3}\right)\sum ^\infty_{n=0}\frac{D_{n}(l_1,l_2,l_3)}{\Gamma\left ( \frac{E}{r} \right )\left ( \frac{E}{r} \right )} \left ( \frac{\bar{\gamma}_{RD}}{m} \right )^{\frac{E}{r}+1} \Upsilon\left (\frac{mT_0}{\bar{\gamma}_{RD}}, u\right ).\label{PER}\hspace{-3.5em}\tag{21}
\end{align}
\noindent\rule{\textwidth}{.5pt}%\vskip3pt
\vspace*{-1.5em}
\end{figure*}
\setcounter{equation}{0}

\begin{figure}[t]
\vspace{-0.5em}
\centering
			\includegraphics[width=0.7\linewidth]{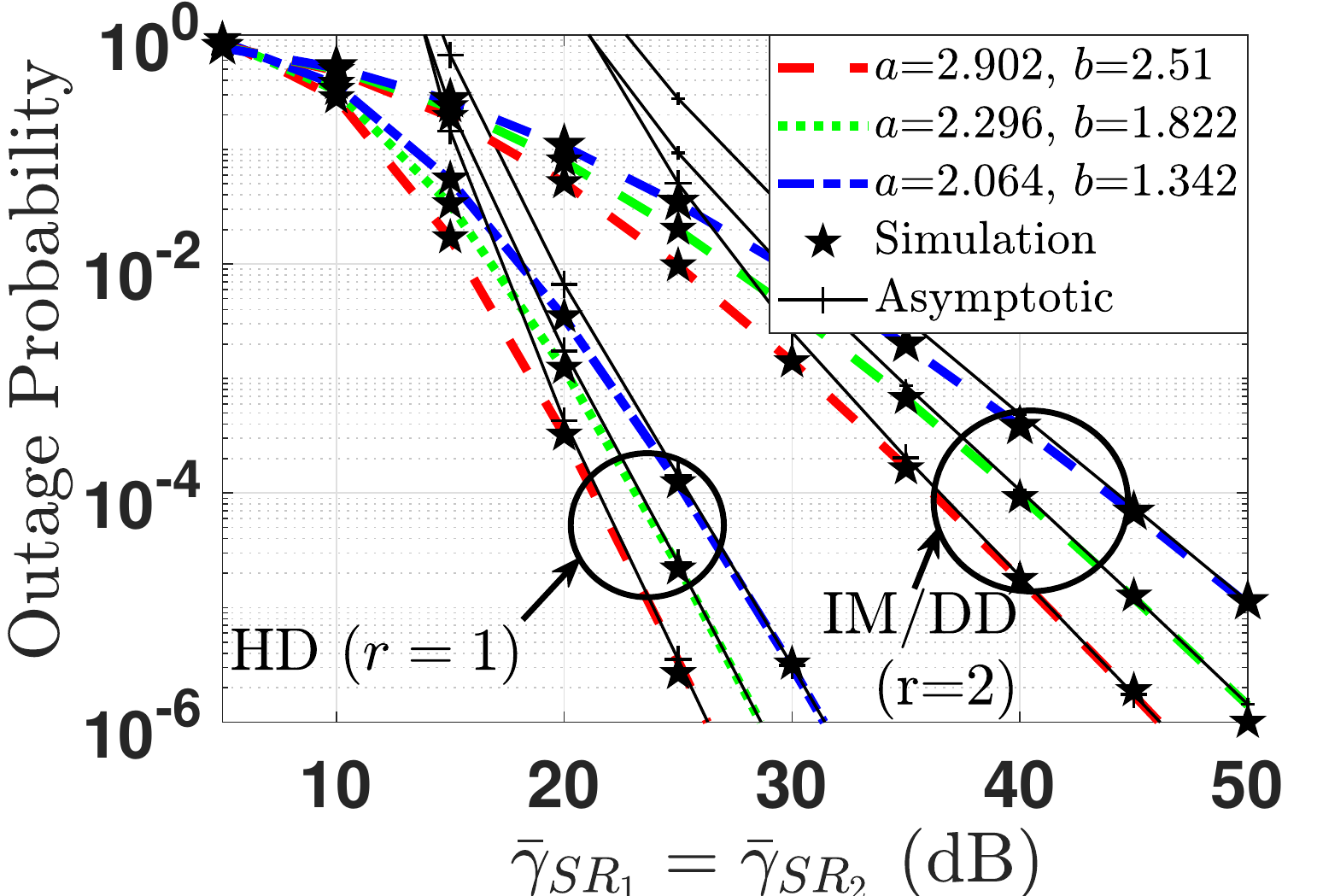}
			\vspace{-0.6em}
										\caption{ Comparison of OP versus SNR for mixed dual-hop FSO-RF system under various AT conditions and detection techniques.}
									\end{figure}
	\vspace{-3.0em}
								\begin{figure}[t]
								\vspace{-1.0em}
								\centering
\includegraphics[width=0.7\linewidth]{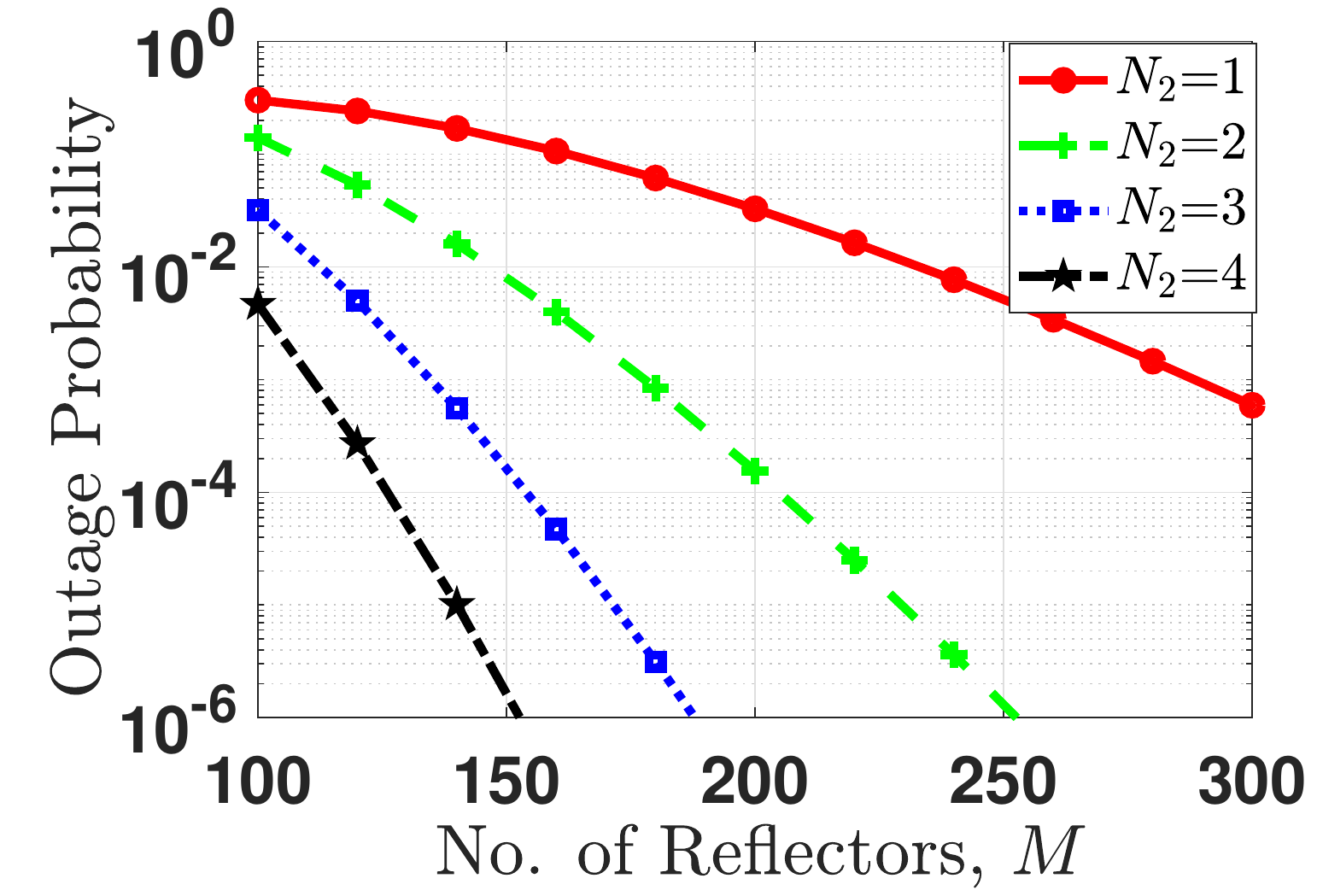}
								\vspace{-0.6em}
									\caption{ Comparison of OP versus number of IRS reflectors for mixed FSO-RF system using H-ARQ with CC
protocol on both links for $N_1$=$3$.}
\vspace{-1.2em}
								\end{figure}
								\begin{figure}[t]
								\vspace{-0.5em}
\centering
\includegraphics[width=0.7\linewidth]{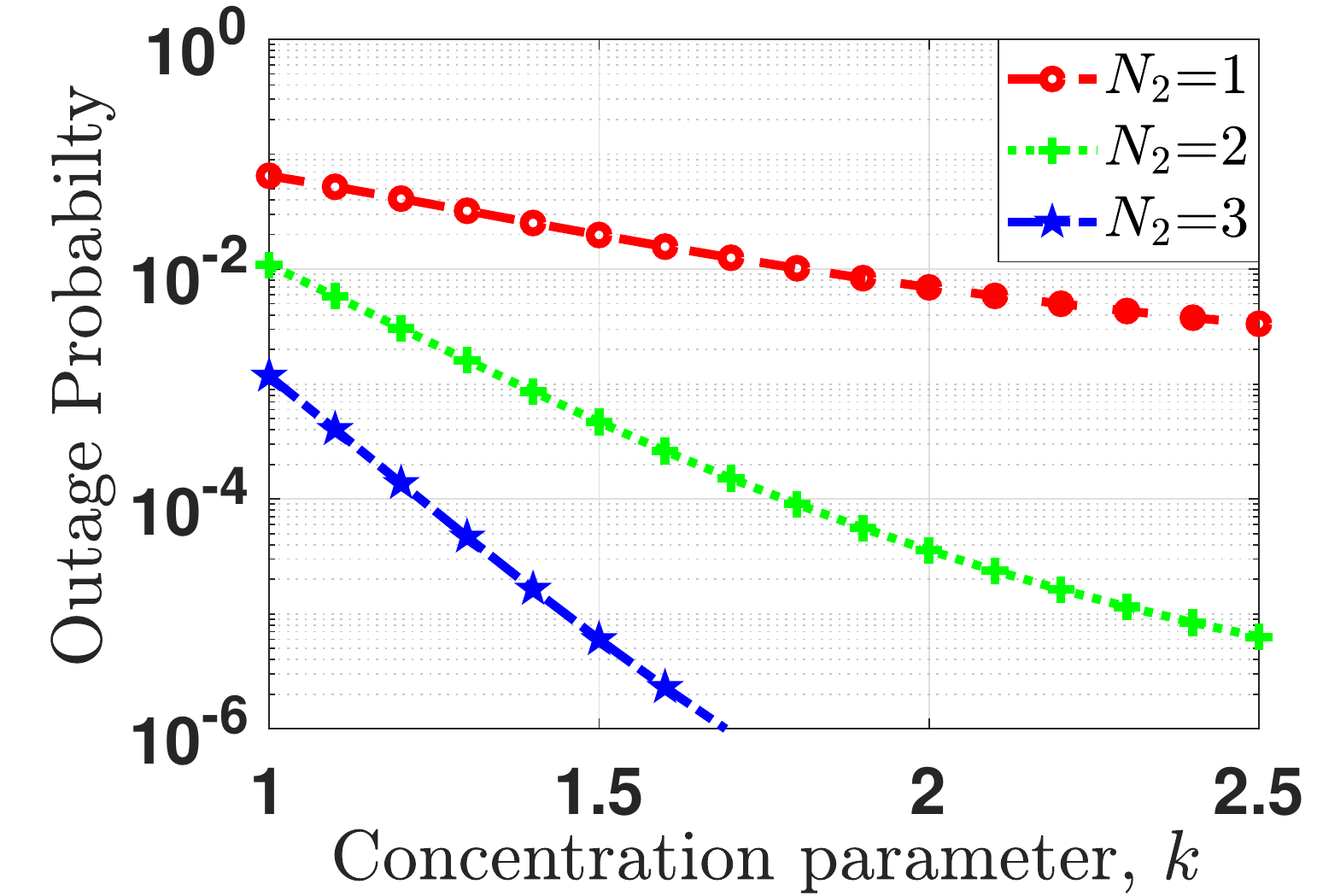}
	\vspace{-0.6em}
										\caption{ OP versus concentration parameter $(k)$ for the various H-ARQ transmission rounds.}
												\vspace{-1em}
							\end{figure}
After some mathematical simplification, Eq. (\ref{I3}) is solved by using \cite[Eq. (2.10.2.2)]{prudnikov:01} as follows: 
\begin{align}
&J_3=\hspace{-6mm}\sum _{{l_1+l_2+l_3}=N_1\atop {0\le l_1,l_2,l_3\le N_1}}\hspace{-0.7em}\left({N_1}\atop {l_1,l_2,l_3}\right)\sum ^\infty _{n=0}\frac{D_{n}(l_1,l_2,l_3)}{\Gamma\left ( \frac{E}{r} \right )\bar{\gamma}_{SR_r}^{\frac{E}{r}}}  \left ( \frac{m}{\bar{\gamma}_{RD}}\right )^{mN_2} \nonumber\\& \times B(1,u)~ _2F_2\left(mN_2+1,u;mN_2+2,u+1;\frac{T_0m}{\bar{\gamma}_{RD}}\right).\label{I3 sol}
\end{align}
In (\ref{I3 sol}), $u$=$E/r+N_2m+1$, $B(.,.)$ is the Beta function \cite[Eq. (8.380)]{gradshteyn:01}, and $_2F_2(.,.;.,.;.)$ is the Hypergeometric function \cite[Eq. (9.14.1)]{gradshteyn:01}. The integral $J_4$ is computed by using (\ref{CDF_SNR}), (\ref{RF_PDF}), and \cite[Eq. (8.31)]{gradshteyn:01} as 
\vspace{-1.5mm}
\begin{align}
 \hspace{-3mm}J_{4}&= \hspace{-6mm}\sum _{{l_1+l_2+l_3}=N_1\atop {0\le l_1,l_2,l_3\le N_1}}\hspace{-0.9em}\left({N_1}\atop {l_1,l_2,l_3}\right)\sum ^\infty_{n=0}\frac{D_{n}(l_1,l_2,l_3)}{\Gamma\left ( \frac{E}{r} \right )\left ( \frac{E}{r} \right )} \nonumber\\& \times \left ( \frac{\bar{\gamma}_{RD}}{m} \right )^{\frac{E}{r}+1}\Upsilon\left (\frac{mT_0}{\bar{\gamma}_{RD}}, u\right ).\tag{20}
\label{I4 sol} 
\end{align}
The derived expressions for $J_i$, $i\in{\left \{ 1,2,3,4 \right \}}$ are substituted into (\ref{PER_integ}) to obtain the PER for the considered mixed FSO-RF communication system and is shown at the top of this page. The value of the waterfall threshold $T_0$ is computed using \cite{ge:01}.   

\subsection{Asymptotic OP Analysis}
 In order to obtain a better understanding of the considered system's performance, we derive the asymptotic expression for the OP. Let us carefully observe (\ref{equivalent_outage_final}) at high SNR. Using the property of the lower incomplete Gamma function that $\Upsilon(z,c)\rightarrow 0$ as $z\rightarrow 0$, the OP of the RF link falls to zero at higher SNR values. As a result, only the $S$-$R$ FSO communication link will dominate at high SNR. Further, in (\ref{equivalent_outage_final}) it is noticed that only the first term in the summation corresponding to $n=0$ will be significant at high SNR. Thus, the asymptotic OP is approximated as
\vspace{-1.50mm}
\begin{equation}
\hspace{-1.5mm}P_{o,CC}\approx \hspace{-4mm}\sum_{{l_1+l_2+l_3}=N_1\atop {0\le l_1,l_2,l_3\le N_1}}\hspace{-2mm}\left({N_1}\atop {l_1,l_2,l_3}\right)\frac{D_{0}(l_1,l_2,l_3)(2^R-1)^{\frac{E}{r}}}{\Gamma\left ( \frac{E}{r} \right )\left ( \frac{E}{r} \right )\bar{\gamma}_{SR_r}^{\frac{E}{r}}}.\vspace{-0.4em} \hspace{-2mm}\label{High SNR CC results}\tag{22}\end{equation}
It is evident from (\ref{High SNR CC results}) that $P_{out,CC}\propto(1/\bar{\gamma}_{SR_{r}})^{\frac{E}{r}}$ at large value of $\bar{\gamma}_{SR_{r}}$ and $E$ consists of $a$, $b$, and $\xi^2$ with possible combinations of $l_1$, $l_2$, and $l_3$ with respect to $N_1$. Thus, the diversity gain is $N_1\min\left\{ \xi^{2}/r,a/r,b/r \right\}$.
\begin{remark}
It is interesting to observe the following: 
\begin{itemize}
\item The diversity of the considered system is dependent only on the FSO link parameters and is independent of the RF link parameters.
\item The number of transmission rounds, detection technique, turbulence parameters, and PE parameter influence the diversity of the considered system.
\end{itemize} 
\end{remark}
\section{Numerical Results and Discussion}
In this section, we will discuss the analytical results presented in the preceding section. Some key system parameter values are $R$=$1$ bps/Hz, $A_{0}$=$1$, $\xi$=$1.2$, $d_{1}$=$d_{2}$=$10$\si{\m}, $M$=$128$, $\nu$=$2.6$, and $(a$=$2.064$; $b$=$1.342)$. Unless otherwise stated, we assume light fog with link visibility and distance of $1$ \si{\km} on the $S$-$R$ link  \cite{verma:02}, \cite{sikri:01}, \cite[Eq. (3.68)]{ghassemlooy:01}. 

Figure 2 shows the OP versus SNR for the considered mixed dual-hop FSO-RF system, where $(a$=$2.064$; $b$=$1.342)$,  $(a$=$2.296$; $b$=$1.822)$, and  $(a$=$2.902$ ; $b$=$2.51)$ represent strong, moderate, and weak AT, respectively. We consider generalized HD ($r$=$1$) and IM/DD ($r$=$2$) techniques on $S$-$R$ hop and H-ARQ with CC protocol transmission rounds $N_1$=$3$ and $N_2$=$2$ with $\bar\gamma_{RD}$=$50$ dB. It can  be observed that the OP improves from strong to weak AT condition. Further, it can be also noticed that the HD technique gives better outage performance compared to IM/DD technique. Moreover, we note that the OPs are $1.562\times10^{-7}$ and  $1.179\times10^{-9}$ at $\bar\gamma_{SR_2}$=$50$ dB and $\bar\gamma_{SR_2}$=$60$ dB, respectively, for $a$=$2.296$, $b$=$1.822$, $\xi$=$1.2$,
\begin{figure}[t]
\vspace{-2.2em}
\centering
\includegraphics[width=0.7\linewidth]{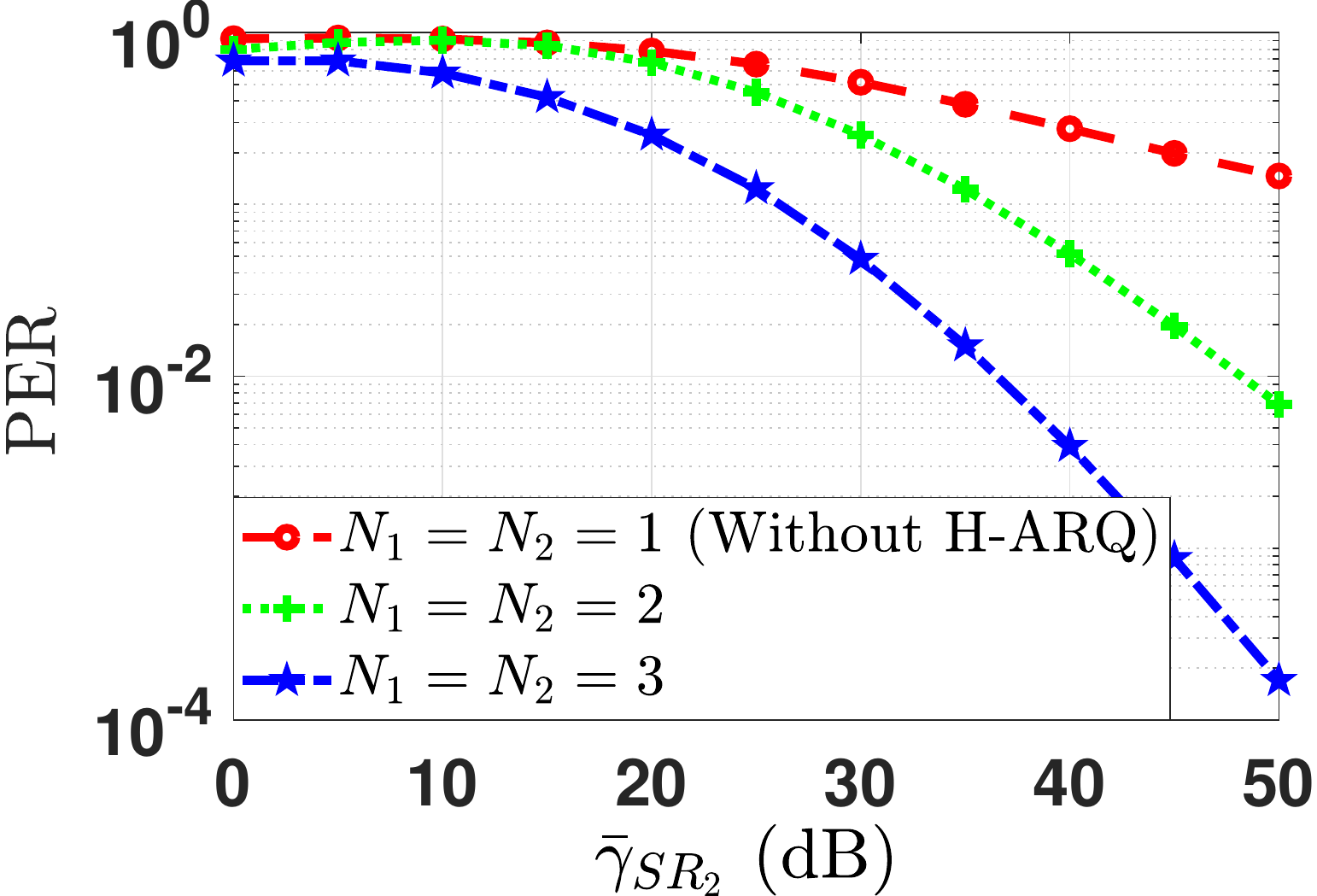}
	\vspace{-0.7em}
								\caption{Comparison of PER versus SNR under strong AT 
										for H-ARQ with CC protocol and IM/DD technique.}
								\vspace{-1.8em}
									\end{figure}
$N_1$=$3$, $N_2$=$2$, and $r$=$2$. Thus, the asymptotic slope is calculated as $\log_{10}(1.562\times10^{-7})-\log_{10}(1.179\times10^{-9})$=$2.12$ $\approx$ $N_1\min\left\{ \xi^{2}/r,a/r,b/r \right\}$=$N_1\xi^2/2$. Likewise, the asymptotic slopes can be calculated and verified for other curves, which validates the asymptotic slope analysis conducted in Section III.D. 

Figure 3 shows the effect of the number of IRS reflectors on the OP of the considered cascaded dual-hop FSO-RF communication system for for $N_1$=3 and various H-ARQ rounds on $R$-$D$ link. We consider $\bar\gamma_{SR_1}$=$45$ dB and $\bar\gamma_{RD}$=$25$ dB on the $S$-$R$ and $R$-$D$ links, respectively. It can be seen that OP performance gets better with the increasing number of reflectors. Moreover, with increasing the H-ARQ transmission rounds, the OP performance improves.

Figure 4 indicates the OP as a function of $k$ for $\bar\gamma_{SR_1}$=$45$ dB and $\bar\gamma_{RD}$=$40$ dB, where parameter $k$ is the concentration parameter of the von Mises distribution which reflects the accuracy of the phase error estimation \cite{badiu:01}. It can be noticed that as parameter $k$ increases, the OP performance enhances. 

The PER versus SNR performance has been plotted for various number of H-ARQ transmission rounds in Figure 5, where we consider the packet length of $1024$ bits and $\bar\gamma_{RD}$=$45$ dB. It can be inferred that PER performance improves with the SNR on the FSO link. Moreover, the PER performance enhances as the H-ARQ transmission rounds increases.

Figure 6 indicates PER versus packet length for different values of $\gamma_{RD}$ and fixed $\bar\gamma_{SR_2}$=$40$ dB. It is observed that as packet length increases, the PER performance slowly degrades and then saturates for higher packet lengths. Moreover, it is also seen that on increasing $\bar\gamma_{RD}$ from $35$ dB to $45$ dB, the PER improves.
	\vspace{-1.5em}
\section{Conclusions}
In this letter, we investigated the performance of a mixed dual-hop FSO-RF system using IRS with phase error on the $R$-$D$ link and H-ARQ protocol over both hops. We derive novel closed-form expressions of OP and PER for the considered system. Useful design insights into the system performance are obtained through the diversity analysis. The presented results comprehensively capture the impact of various system parameters such as AT, number of
	\begin{figure}[t]
\vspace{-2.2em}
\centering
\includegraphics[width=0.7\linewidth]{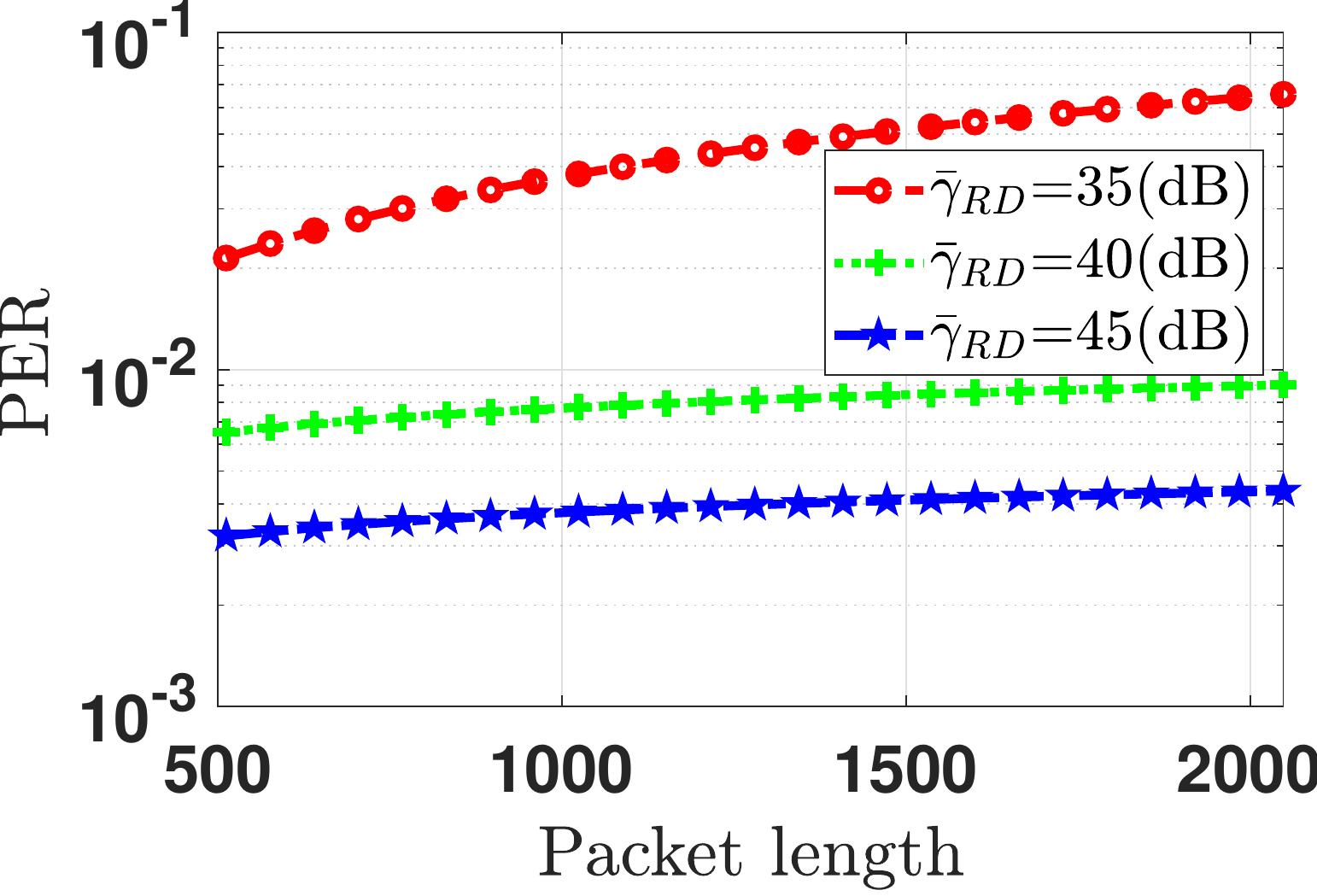}
	\vspace{-0.7em}
										\caption{Comparison of PER versus packet length under strong AT  for H-ARQ with CC protocol and IM/DD technique for $\gamma_{SR_2}$=40 dB.}
										\vspace{-1.5em}
									\end{figure}
IRS reflectors, concentration parameter, H-ARQ rounds, and detection techniques on the outage and PER that are critical for the design of such mixed FSO-RF systems.

\bibliographystyle{IEEEtran}
\bibliography{ref01}

\begin{thebibliography}{10}
\providecommand{\url}[1]{#1}
\csname url@samestyle\endcsname
\providecommand{\newblock}{\relax}
\providecommand{\bibinfo}[2]{#2}
\providecommand{\BIBentrySTDinterwordspacing}{\spaceskip=0pt\relax}
\providecommand{\BIBentryALTinterwordstretchfactor}{4}
\providecommand{\BIBentryALTinterwordspacing}{\spaceskip=\fontdimen2\font plus
\BIBentryALTinterwordstretchfactor\fontdimen3\font minus
  \fontdimen4\font\relax}
\providecommand{\BIBforeignlanguage}[2]{{%
\expandafter\ifx\csname l@#1\endcsname\relax
\typeout{** WARNING: IEEEtran.bst: No hyphenation pattern has been}%
\typeout{** loaded for the language `#1'. Using the pattern for}%
\typeout{** the default language instead.}%
\else
\language=\csname l@#1\endcsname
\fi
#2}}
\providecommand{\BIBdecl}{\relax}
\BIBdecl

\bibitem{khaligi:01}
M.~A. {Khalighi {\em et al.}}, ``Survey on free space optical communication: A
  communication theory perspective,'' \emph{IEEE Commun. Surveys Tuts.},
  vol.~16, no.~4, pp. 2231--2258, Fourthquarter 2014.

\bibitem{yang:02}
F.~{Yang {\em et al.}}, ``Free-space optical communication with nonzero
  boresight pointing errors,'' \emph{IEEE Trans. Commun.}, vol.~62, no.~2, pp.
  713--725, February 2014.

\bibitem{verma:01}
G.~D. Verma~{\em et al.}, ``Secrecy performance of {FSO} communication systems
  with non-zero boresight pointing errors,'' \emph{IET Commun.}, vol.~15,
  no.~1, pp. 155--162, 2021.

\bibitem{yang:01}
L.~Yang~{\em et al.}, ``Mixed dual-hop {FSO-RF} communication systems through
  reconfigurable intelligent surface,'' \emph{IEEE Commun. Lett.}, vol.~24,
  no.~7, pp. 1558--1562, 2020.

\bibitem{3gpp:01}
``{3GPP TR38.885 V16.0.0 (38885-g00.zip), Study on NR Vehicle-to\\-Everything
  (V2X)},
  {\url{https://www.3gpp.org/ftp/specs/archive/38\_series\\/38.885}},{Accessed:
  18-07-2021}.''

\bibitem{mathur:01}
A.~Mathur~{\em et al.}, ``Performance of hybrid {ARQ} over power line
  communications channels,'' in \emph{2020 IEEE 91st . Veh. Technol. Conference
  (VTC2020-Spring)}, 2020, pp. 1--6.

\bibitem{ge:01}
S.~Ge~{\em et al.}, ``Packet error rate analysis and power allocation for
  {CC-HARQ} over {Rayleigh} fading channels,'' \emph{IEEE Commun. Lett.},
  vol.~18, no.~8, pp. 1467--1470, 2014.

\bibitem{xi:01}
Y.~Xi~{\em et al.}, ``A general upper bound to evaluate packet error rate over
  quasi-static fading channels,'' \emph{IEEE Trans. Wireless Commun.}, vol.~10,
  no.~5, pp. 1373--1377, 2011.

\bibitem{verma:02}
G.~D. Verma~{\em et al.}, ``Performance improvement of {FSO} communication
  systems using hybrid-{ARQ} protocols,'' \emph{Appl. Opt.}, vol.~60, no.~19,
  pp. 5553--5563, Jul 2021.

\bibitem{bjornson:01}
E.~Bjornson~{\em et al.}, ``Intelligent reflecting surface versus
  decode-and-forward: How large surfaces are needed to beat relaying?''
  \emph{IEEE Wireless Commun. Lett.}, vol.~9, no.~2, pp. 244--248, 2020.

\bibitem{sikri:01}
A.~Sikri~{\em et al.}, ``Reconfigurable intelligent surface for mixed {FSO-RF}
  systems with co-channel interference,'' \emph{IEEE Commun. Lett.}, vol.~25,
  no.~5, pp. 1605--1609, 2021.

\bibitem{ai:01}
Y.~Ai~{\em et al.}, ``On hybrid-{ARQ}-based intelligent reflecting
  surface-assisted communication system,'' \emph{arXiv preprint
  arXiv:2009.10776}, 2020.

\bibitem{jamali:01}
V.~Jamali~{\em et al.}, ``Intelligent surface-aided transmitter architectures
  for millimeter-wave ultra massive {MIMO} systems,'' \emph{IEEE Open J.
  Commun. Soc.}, vol.~2, pp. 144--167, 2021.

\bibitem{badiu:01}
M.-A. Badiu~{\em et al.}, ``Communication through a large reflecting surface
  with phase errors,'' \emph{IEEE Wireless Commun. Lett.}, vol.~9, no.~2, pp.
  184--188, 2020.

\bibitem{ghassemlooy:01}
Z.~Ghassemlooy~{\em et al.}, \emph{Optical wireless communications: system and
  channel modelling with Matlab{\textregistered}}.\hskip 1em plus 0.5em minus
  0.4em\relax CRC press, 2019.

\bibitem{bhatnagar:01}
M.~R. Bhatnagar~{\em et al.}, ``Performance analysis of {Gamma–Gamma} fading
  {FSO} {MIMO} links with pointing errors,'' \emph{J. Lightw. Technol.},
  vol.~34, no.~9, pp. 2158--2169, 2016.

\bibitem{gradshteyn:01}
I.~S. Gradshteyn~{\em et al.}, \emph{Table of integrals, series, and products},
  7th~ed.\hskip 1em plus 0.5em minus 0.4em\relax Elsevier/Academic Press,
  Amsterdam, 2007.

\bibitem{kong:01}
L.~Kong~{\em et al.}, ``Effective rate evaluation of {RIS}-assisted
  communications using the sums of cascaded $\alpha$-$\mu$ random variates,''
  \emph{IEEE Access}, vol.~9, pp. 5832--5844, 2021.

\bibitem{gupta:01}
A.~Gupta~{\em et al.}, ``Cascaded {FSO-VLC} communication system,'' \emph{IEEE
  Wireless Commun. Lett.}, vol.~6, no.~6, pp. 810--813, 2017.

\bibitem{prudnikov:01}
A.~P. Prudnikov~{\em et al.}, \emph{Integrals and Series of Special Functions.
  Vol.2}.\hskip 1em plus 0.5em minus 0.4em\relax RUS: Science, 1983.

\end{thebibliography}

\end{document}